\documentclass[a4paper,11pt]{article}

\usepackage{adjustbox}
\usepackage[utf8]{inputenc}
\usepackage{amsmath}

\usepackage{adjustbox}

\begin{document}

\title{Proposal for the resolution of the disagreement between DAMA and other
  Underground Dark Matter Experiments }
\author{C. D. Froggatt, Glasgow University\\
H.B.Nielsen, Niels Bohr Institute}

\maketitle


\begin{abstract}
  It is a great mystery still needing a solution that the underground
  experiments looking for dark matter disagree, in the sense that DAMA/LIBRA
  has found dark matter with a confidence level of several standard deviations, while neither the
  xenon based experiments nor even the experiments using the same target material NaI(Tl) as DAMA 
  see the same amount of dark matter. We propose to solve this enigma by
  postulating that the dark matter gets excited during its passage through
  the mountain shielding from outer space down to the experimental halls; the DAMA experiment is namely about twice as deep as the ANAIS and COSINE-100
  experiments, so that there is longer passage time for DAMA, which thus gets
  more excited dark matter particles. The xenon-scintillator experiments
  have their strong limits for dark matter making itself felt via nucleus
  recoils, and thus we can just assume that the excited dark matter
  mainly emits e.g. electrons, and the xenon-experiments would hardly
  see such dark matter. The single event recently seen by LUX/ZEPPELIN could be
  due to a low rate of nucleus emission in addition to the
  hypothesised electrons.
\end{abstract}
\section{Introduction}
It is a great mystery that the DAMA/LIBRA underground experiment \cite{DAMA18,
  DAMA22, DAMA25} looking to
directly  observe dark matter causing scintillation in NaI (doped with
Tl) seemingly sees the seasonal oscillations in the counting rate expected
for dark matter with a statistical significance of 13.7 standard deviations, while all other such
underground experiments \cite{LUX, XENON, ANAIS, COSINE100, combined} also looking for dark matter get null results.

In the present article we propose a possible solution to this long-standing problem as being due to the fact that the DAMA/LIBRA experiment is about
twice as deep under the Earth surface as two other experiments using the
same scintillation technique as DAMA/LIBRA itself, namely ANAIS and COSINE-100.
So, assuming that the dark matter gets excited on its way through the
shielding earth layers, it will get more excited when reaching DAMA/LIBRA
than when reaching the other experiments. It is also assumed that
the dark matter loses most of its velocity on its way through the earth
shielding and emits its excitation energy mainly as say electrons or X-ray
photons. This assumption that the radiation seen by DAMA/LIBRA
is not of the usually assumed type of recoil nuclei, but rather of the type
of electrons say being emitted, immediately makes the chances of observation by the many xenon-scintillator based underground experiments, like LUX and XENONnT, become much less than if it had been nuclei which DAMA/LIBRA had seen.

Such a model as we propose may seem a bit too complicated to be immediately
convincing, but it should be borne in mind that none of the competing ideas
for explaining the mystery is so attractive. 
Researchers in the dark matter community have suggested that the DAMA/LIBRA-signal is probably due to some, as yet unidentified, 
uncompensated season-conditioned changes in the laboratory or instrumental
noise (such as unstable background conditions or data-treatment) rather
than true particles \cite{Maruyama}. The upcoming experiments COSINUS \cite{cosinus} and SABRE \cite{sabre} with NaI(Tl) detectors at the same Gran Sasso National Laboratory as DAMA will provide important checks of the DAMA/LIBRA signal.

In the present work we propose that dark matter has properties
very different from the usually believed WIMP type of model. Our
proposal namely deviates from this usual picture by:

\begin{itemize}
\item{\bf Interacting}
  The crucial point needed to solve the mystery is that the dark
  matter interacts with ordinary matter so as to get excited from this
  interaction. But once one believes in such an interaction it is suggestive
  - but of course not strictly needed - that the dark matter may partly or
  fully have to consist of ordinary matter in some modified form, and thus
  also a significant self-interaction would be natural.

  The interaction should also be strong enough to bring down the galactic speed
  of the dark matter particles so they can no longer cause the nuclear recoils
  which the experiments like LUX and XENONnT effectively look for.
  In this way it should explain why experiments definitively looking
  for nuclear recoils coming from galactic speed particles do not find any.
\item{\bf Excitable}
  For the solution of the mystery we propose, it is crucial that the
  dark matter has excited states, so that excitation in the earth shielding
  can take place.
\item{\bf Radiating} While physicists usually expect that the
  observational signal for direct dark matter should be a recoil due to
  the fast moving dark matter particles (with galactic velocity), we propose that
  the signal is due to emission of some radiation from the
  excited dark matter.

  We have in mind that this radiation shall mainly consist of electrons,
  thereby making it very difficult to get counted as a signal for dark matter by the xenon-based
  experiments. They namely typically concentrate on seeing the nuclear
  recoil  events and keeping them separate from the background.
  \end{itemize}

The paper is organized as follows:

In the following section \ref{assumptions} we put forward the assumptions
about the properties of the dark matter constituents which are crucial
for our way of circumventing the seemingly paradoxical disagreements 
of the various underground experiments. In section \ref{happens} we then
describe what, according to our type of model, is supposed to happen when the
dark matter meets the Earth. In section \ref{discussion} we seek to outline which of these assumptions are truly needed and therefore have to be upheld in the final model for dark matter.
We have a special section \ref{LZn} about the single event of potentially
a genuine WIMP-nucleus-recoil (anyway still far too weak a signal to agree with DAMA, if it were a WIMP).
In section \ref{conclusion} we
conclude by again emphasizing that, in the light of the not very convincing
alternatives for explaining the ``mystery'', one is almost driven to
a model of the type we propose.

\section{Our Basic Assumptions}\label{assumptions}
The basic assumptions, which we have to make about the dark matter in order
to  at least crudely solve the disagreement problem, are not really so terribly
strong, although they of course specify a special class of dark matter
models. Indeed we shall argue that the needed assumptions can be
achieved in the model with which we have worked on for many years, and in which
the dark matter consists of \textquotedblleft pearls" of ordinary matter held under high pressure
by some skin or domain wall separating different phases of the vacuum \cite{hemispheres}. There is even the
possibility that the hoped for different phases of the vacuum could come out
of strong interaction physics and be seen in so-called Columbia plots \cite{columbia,corfu2022} of lattice QCD.

It should be possible to construct other dark matter models in this special class, which differ appreciably from our model and are also able
to resolve the disagreement between DAMA and the other underground experiments.
The crux of the matter for such models are the following assumptions:

\begin{itemize}
\item{\bf 1. Excitation during passage} {\em The dark matter particles shall be
  excitable and be excited during the passage through the earth / mountain
  shielding above the experiments.}

  It is not important whether this excitation uses the kinetic energy of
  the dark matter particle, i.e. an inelastic scattering takes place,
  or it is some interaction taking place in which the nuclei or the electrons
  in the shielding get absorbed and bound so as to provide energy for heating
  or excitation.

\item{\bf 2. Smoothness of lifetime distribution} {\em The excitations 
  shall decay with various lifetimes, and it shall be so that
  the probability distribution of lifetimes of the order of the passage
  time through the earth shielding counted on a logarithmic axis shall be
  smooth / slowly varying, and sufficiently strong that radiation
from nearby interactions do not take over.}

  It should be noted, that we do not need to assume that lifetimes of the
  order of the passage time dominate the decays in general. It is enough
  that there are
  some possibilities for lifetimes of this order and that the variation
  of the probability for various lifetimes in that region varies slowly,
  and that the emission from short lifetime decays excited very near the
  counters in the experiments do not take over.

\item{\bf 3. Decay via electrons or photons} {\em It is important that
  the major part of the decays do not emit nuclei, because the xenon
  scintillator experiments  most effectively look for recoil nuclei, and
  would have seen the decays if there were a significant branching ratio into nuclei}

  In fact if we assume the dominant decays are into electrons as the emitted
  particle, we expect the dark matter particle to be loaded up with positive
  charge after emission of several electrons, and for keeping it approximately neutral it would need to emit e.g. a nucleus from time to time.
  But since nuclei could have appreciably higher charge than an electron such
  nucleus emission could be much more seldom. (We think the single
  nucleus ``recoil'' event seen by LUX/ZEPPELIN \cite{LUXnew} could be such a rare nucleus emission.)
\end{itemize}

\section{What happens then}
\label{happens}
Under the assumptions about the properties of dark matter in the foregoing
section \ref{assumptions}, the dark matter particles will now get excited
as they pass down through the earth shielding above the experimental hall. As suggested by the assumptions 1. and 2. the excitations are distributed according
to lifetime. The excitations with lifetime shorter than or of the order of the
passage time will decay in the shielding mountain relatively quickly after the
excitation. The excitations with much longer lifetimes than the passage time
will first decay deeply after the experimental hall was passed. Only the
excitations with lifetimes of a similar order of magnitude as the passage
time will have a good chance to decay in the experimental detectors.

But now the passage time during which the excitations are made corresponds
to a depth of 1400 m in the case of DAMA, while COSINE-100 is at 700 m
and ANAIS at 850 m. 
Thus we can crudely say that the dark matter particles reaching DAMA have gotten about twice as long a distance in which to get excited. If the relevant
lifetime is of the order of the passage time - or preferably even longer -
the degree of excitations of the particles decaying in DAMA will be about
twice as big as in ANAIS and COSINE-100.

Our predictions for the amplitudes of dark matter annual modulation $S_{m\; pred.}$ at COSINE-100 and ANAIS are then given in terms of the amplitudes  $S_{mDAMA}$ reported by DAMA/LIBRA \cite{DAMA18}:
\begin{eqnarray}
S{mCosine100\; pred.} & =  &\frac{700}{1400}S_{mDAMA} \\
S{mAnais\; pred.} & =  &\frac{850}{1400}S_{mDAMA}
\end{eqnarray}
The values of the DAMA/LIBRA amplitudes and of our predicted amplitudes for COSINE-100 and ANAIS are listed below together with the experimental values $S{m\; exp}$ obtained by COSINE-100 \cite{COSINE100} and ANAIS \cite{ANAIS}. The deviations between our predictions and the experimental values are also given and compared to the deviations obtained using the DAMA/LIBRA amplitudes, as usually done before, rather than our predicted amplitudes. As can be seen, the deviations for our model predictions are now all less than 3 standard deviations and the model is consistent with experiment.
  
\begin{eqnarray}
	\hbox{\bf DAMA/LIBRA:} &&\nonumber\\
  \hbox{For 1-6 kev region:} S_{mDAMA} & = & 0.0104\pm 0.0007 cdp/kg/keV\\
  \hbox{for 2-6 kev region:} S_{mDAMA} &=& 0.01014\pm 0.00074 cdp/kg/keV\\
  \hbox{\bf Predict for COSINE-100:} &&\nonumber\\
    \hbox{for 1-6keV region:} S_{mCosine100\; pred.} &= & 0.0052\pm 0.00035
    cdp/kg/keV\\
    \hbox{to compare exp: } S_{mCosine100 \; exp} &=& 0.0017\pm 0.0029 cpd/kg/keV\\
    \hbox{i.e. } deviation &=&{ \bf 1.2 s.d.}\\
    \hbox{while before } deviation_{before} &=& 3.0 s.d.\\
    \hbox{for 2-6 keV region:} S_{mCosine100\; pred.}&=&
    0.00507\pm 0.00037 cpd/kg/keV\\
    \hbox{to compare exp: } S_{mCosine100 \; exp} &=& 0.0053 \pm 0.0031 cpd/kg/keV\\
    \hbox{i.e. } deviation &=& {\bf 0.07 s.d.}\\
    \hbox{while before: } deviation_{before}&=& 1.6 s.d.\\
    \hbox{\bf  Predict for ANAIS :} && \nonumber\\
    \hbox{for 1-6 keV region: } S_{mANAIS\; pred.}&=&0.0063\pm 0.0004 cpd/kg/keV\\
    \hbox{to compare exp: } S_{mAnais\;  exp}&=&-0.0004\pm 0.0025 cpd/kg/keV\\
    \hbox{i.e. } deviation &=& {\bf 2.7 s.d.}\\
    \hbox{while before: } deviation_{before} &=& 4.3 s.d.\\
    \hbox{for 2-6 keV region: }S_{mANAIS \; pred.}&=&0.0062\pm 0.00045
    cpd/kg/keV\\
    \hbox{to compare exp: } S_{mANAIS\; exp}&=& 0.0011\pm 0.0025 cpd/kg/keV\\
    \hbox{i.e. } deviation &=& {\bf 2.0 s.d.}\\
    \hbox{while before: } deviation_{before}&=& 3.6 s.d.
  \end{eqnarray}

\section{Discussion}
\label{discussion}
We would now like to argue that we are to some extent driven to a picture much
like the one just described of dark matter being excited and radiating
presumably electrons or occasionally photons too:

\begin{itemize}
\item{\bf 1. Not recoil nuclei} The xenon-scintillator
  experiments, such as LUX/ZEPPELIN, do not at all see the intensity
  of dark matter impacts as seen by DAMA/LIBRA. This suggests that what DAMA/LIBRA
  sees is {\bf not} recoil nuclei as expected in WIMP theories of dark
  matter, but rather e.g. electrons, for which there is a huge background
  and which would mainly just be considered as part of the  background by the xenon-experiments.

\item{\bf 2. Not recoil at all} If the dark matter particles were
  still neutral - as dark matter at least must be out in space - when
  reaching the underground experiment, then they could hardly avoid giving
  most energetic pushes to the nuclei rather than to the electrons, except if
  they had a much stronger coupling to electrons than to nuclei.
 Unless the dark matter particles interact with the earth shielding
  on the way down, they must still be neutral when reaching the
  experiments and electron recoil dominance is disfavoured.
 So rather an emission of the electrons from the dark matter is
  suggested.

\item{\bf 3. Earth interaction needed} An emission of something
  from dark matter is of course imaginable as just a generally unstable
  dark matter particle decaying continuously. But if it should decay
  often enough to be seen in DAMA of course the decay time would be so
  short that all dark matter in the Universe would have decayed long ago.
  So decaying  dark matter is only possible, if the dark matter is getting
  excited during its passage through the earth.
\item {\bf 4. Excitation or slow down} The effect on a dark matter particle
  passing through the earth shielding above one of the underground experiments
  could be a slow down and/or it could be an excitation of internal
  degrees of freedom. If the particle remains neutral a slow down
  would mean that the number of collisions would still be given by the
  cross section and the number of particles, but the energy of the recoil
  would be lower the deeper the particle comes. The spectrum of the events
  should thus be pushed lower and lower in energy the deeper the experiment
  is situated. If some signals would hereby pass under the energy
  limit for observation around 1 keV or 2 keV a lowering of energy would mean
  a diminishing of the signal. This effect would go the wrong way, in the sense
  that it would lead to fewer dark matter events in DAMA/LIBRA which is
  deeper and would get slower particles than in ANAIS and
  COSINE-100.

  On the other hand a longer path with excitation should lead to a more excited
  particle emitting a stronger signal. So with excitation being the important effect we would get the right sign of the difference in counting rates in DAMA/LIBRA versus the higher up experiments.

  If the dark matter particle can get charged by passing through the earth
  shielding, one might imagine it delivering a larger
  stopping energy as the particle gets slower (the Bragg peak).
  But of course getting an a priori neutral dark matter particle charged
  is really an excitation. So in any case excitation of some sort is
  favoured.

\item{\bf 5. Excitation lifetime as sinking time} It is important that 
  an appreciable part of the DAMA /LIBRA signal should come from the excitations
  made at a height of more than 700 m above the detector, and really in the upper part of the 1400 m of earth shielding. Otherwise the decays of the excitations or their observation somehow in DAMA would not  
  be seen at different rates from those at the higher up experiments ANAIS and COSINE-100. So it is needed that preferably the dominant
  part of the DAMA-signal comes from decays or observations of excitations
  with - at least - a lifetime similar to the time for the sinking down
  or passage time through the shielding.

  It does not mean that there could not be a perhaps much more copious
  decay of shorter living excitations, provided they for some reason
  are not accepted as single emission events. If the dominant lifetime
  of the excitations is much shorter than the sinking time, it could be that
  the dark matter particle emits so many say electrons in a short time
  that, if it happens in the detector, it becomes a multiple hit event.
  
  Perhaps the easiest way to achieve the required scenario is for the life-times of the various excitations to have a smooth distribution in the logarithm. Such a smooth distribution in the logarithm might be obtained if a tunnelling effect is needed for the decays of the excitations. This suggests that there are sufficiently long and or strong tunnelling possibilities in the dark matter material to give long decay times of the order of the sinking time.

  If we do not have sufficiently long decay times we see that the
  separation of the predicted rates for the experiments at different depths is not possible. So we can say it is crucial for understanding the mysterious much higher rate of the modulation amplitude for DAMA than for ANAIS and COSINE-100, that we have such relatively long lifetime excitations of the dark matter particles.

\item{\bf 6. More than collision energy} The energies per event seen
  by DAMA/LIBRA is in the range from as low as observable 1 to 2 keV up to
  6 keV. Electrons of this energy can hardly be just produced by galactic
  velocity particles hitting electrons. It might be produced by collision
  accelerated nuclei, but that would have been seen by xenon experiments like
  LUX/ZEPPELIN. So we are driven to think that the energy for these
  electrons rather comes from some reaction between the dark matter particles
  and the atoms in the shielding above the experiments.
  It means that electrons or nuclei should be able to bind to the dark
  matter with a significant release of binding energy, enough to produce
  in turn the electrons with a few keV energy.

  The particles reaching down to the experiments would thus likely be bound
  states of the original dark matter particles and atoms or parts of atoms
  absorbed from the earth shielding. At least after the first absorption the dark matter particle - now
  excited - would contain electrons and/or nuclei.

  This point of absorption energy being needed really calls for models
  like  our own of ordinary matter compressed by domain walls  \cite{hemispheres} or the composite XHe dark atom model of Khlopov and collaborators \cite{khlopov}.
  The binding energies must then at least be enough to excite electrons to
  a few keV's.
  
  \item{\bf 7. Southern Hemisphere} With such a strong interaction of the dark
  matter as we imagine in this article, the dark matter has no chance to pass
  through the Earth, rather only  sinking due to gravity, maybe into the
  centre of the Earth. So no dark matter will, like
  in WIMP theories, go in at one place and out on the other side of the Earth.
  Rather the laboratories on one side of the Earth can only hope to detect
  particles coming from that side. Now the motion of the Earth and Sun relative
  to the dark matter is in the direction towards the star Deneb, which is
  in the Northern Hemisphere. So the Southern Hemisphere of the Earth
  is so to speak being hit from the backside. This actually means that the
  seasonal modulation on the Southern hemisphere shall be the opposite
  of the modulation on the Northern Hemisphere. So, in our type of models, we are
  predicting that the experiments in the Southern Hemisphere shall
  have the modulation of their dark matter with a half year shifted phase
  compared to the one on the Northern Hemisphere \cite{hemispheres}.
  
  From continuity of the effect to be seen around the Earth we can conclude
  that there will be no modulation at the equator and maximal modulation on the
  poles. Thus, in order to get a stronger signal, it would be advisable to build
  an experiment closer to one of the poles. 
  \end{itemize}

\section{The Seemingly Nuclear Recoil in LUX/Zeppelin}
\label{LZn}
Of course the recent single seemingly WIMP-like dark matter event seen by
LUX/ZEPPELIN \cite{LUXnew} is not at a sufficient event rate to bring consistency with the 
DAMA/LIBRA
results in a WIMP model. So we would anyway need some other explanation than
a simple WIMP model to cope with both the LUX/ZEPPELIN and the DAMA/LIBRA results, such
as the one in the present article, say. Then the question is if our
type of dark matter could produce the recent particle seen in LUX/ZEPPELIN.

For this we propose that indeed the dark matter particles  may have to emit
something positively charged from time to time, in order not to end up with a high positive charge. In fact our suggestion above was that the radiation seen
by DAMA/LIBRA was {\em electrons} emitted from excited dark matter
particles. If a dark matter particle emits many electrons it becomes
of course positively charged, and at some point it might become so highly
charged that it begins to emit nuclei. In order for the dark matter
particles to keep approximately neutral in the long run, they have to emit
nuclei and electrons in amounts inversely proportional to the charges
of these emitted particles. Thinking e.g. of an iron nucleus with $Z=26$,
there would come about one nucleus for each 26 electrons in the full radiation.


Thus even if it should turn out that the  recently observed
particle at LUX/ZEPPELIN should be a dark matter particle, for example by such
particles being observed to have the characteristic dark matter
seasonal modulation, our picture could be upheld.

\section{Conclusion}\label{conclusion}

We have here proposed a couple of properties of the dark matter,
which could at least roughly resolve the paradox or mystery, that the
DAMA/LIBRA direct dark matter seeking experiment gets results in disagreement
with all the other similar underground experiments. First we suggest that the disagreement between DAMA/LIBRA and the other NaI(Tl) based experiments is
understandable as due to DAMA/LIBRA being placed crudely twice as deeply
as ANAIS  and COSINE-100. Next the deviation between
the DAMA/LIBRA results and the xenon-based experiments LUX/ZEPPELIN,
XENONnT, etc. is due to the fact that the latter experiments mainly look for
nuclear recoils, while we suggest that what DAMA/LIBRA saw was electrons
radiated out of the excited dark matter particles. The point of the dependence on the depth under the Earth surface is in our suggestion, that
the excitation of the dark matter causing it to radiate was brought about
during the passage through the earth shielding above the experimental hall.
Thus the deeper the experiment is placed the more the excitation.

We suggested that if our picture should be true, then a bigger rate of
dark matter observation than even that of DAMA/LIBRA should be possible by going even deeper than DAMA's 1400 m, or even closer to one of the poles
than DAMA.

Our type of model also has the property that the effect of seasonal modulation gets the opposite sign on the Southern Hemisphere.

\end{document}